# Label Adversarial Learning for Skeleton-level to Pixel-level Adjustable Vessel Segmentation

Mingchao Li[1], Kun Huang[1], Zetian Zhang[1], Xiao Ma[1] and Qiang Chen[1]

[1] School of Computer Science and Engineering, Nanjing University of Science and Technology, Nanjing 210094, China
Chen2qiang@njust.edu.cn

**Abstract.** *You can have your cake and eat it too.* Microvessel segmentation in optical coherence tomography angiography (OCTA) images remains challenging. Skeleton-level segmentation shows clear topology but without diameter information, while pixel-level segmentation shows a clear caliber but low topology. To close this gap, we propose a novel label adversarial learning (LAL) for skeleton-level to pixel-level adjustable vessel segmentation. LAL mainly consists of two designs: a label adversarial loss and an embeddable adjustment layer. The label adversarial loss establishes an adversarial relationship between the two label supervisions, while the adjustment layer adjusts the network parameters to match the different adversarial weights. Such a design can efficiently capture the variation between the two supervisions, making the segmentation continuous and tunable. This continuous process allows us to recommend high-quality vessel segmentation with clear caliber and topology. Experimental results show that our results outperform manual annotations of current public datasets and conventional filtering effects. Furthermore, such a continuous process can also be used to generate an uncertainty map representing weak vessel boundaries and noise.

## 1 Introduction

Optical coherence tomography angiography (OCTA) is a noninvasive imaging modality that allows the visualization of retinal microvessels with micron-level resolution, thus being increasingly used in the analysis of various retinal vascular diseases [1]. Quantitative analysis of en-face OCTA images (Fig. 1a) usually requires the segmentation of the retinal vessels and capillaries. Different vessel segmentation methods have been explored in recent years, including filter-based methods [2-5] and supervised-based methods [6-10]. In particular, supervised-based deep learning methods have drawn a lot of attention because of their excellent segmentation performance. They usually require manual annotations that can be regarded as ground truth. However, manually drawing large vessels in OCTA images has been extremely time-consuming, while manually drawing capillaries is even more exhausting due to their dense structure. Noise, poor contrast and low resolution aggravate the difficulty. For these reasons, existing public datasets (such as OCTA-500 [11], ROSE [10]) mainly focus on large vessels with limited capillaries. To date, microvessel segmentation with high-detailed capillaries in OCTA images is still a challenging task.

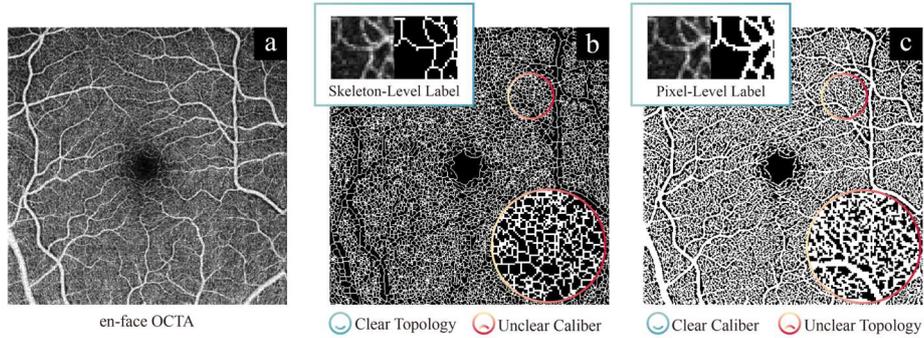

**Fig. 1.** Preliminary experiments on vessel segmentation in OCTA images. (a) An example of en-face OCTA image. (b) Skeleton-level segmentation. (c) Pixel-level segmentation.

In this work, we try to segment full vessels (including large vessels and capillaries) in OCTA images. To this end, we made two types of labels (skeleton-level labels and pixel-level labels, as shown in Fig. 1, blue box). They are commonly used in supervising vessel segmentation [10]. Using them to train convolution neural networks (e.g., U-Net [12]) can achieve skeleton-level segmentation (Fig. 1b) and pixel-level segmentation (Fig. 1c). The skeleton-level segmentation shows a clear topology of capillaries, but it ignores the vessel diameter information, thus it is difficult to distinguish between large vessels (arteries and veins) and capillaries. The pixel-level segmentation shows a clear vessel caliber, but its topology is unclear when the vessels are dense and with noise, artifacts and motion blur. The strengths and weaknesses of the two segmentations are complementary, and we thus guess whether there exists an intermediate state with the advantages of both.

To close this gap, we propose a novel deep learning approach called Label Adversarial Learning (LAL) for skeleton-level to pixel-level adjustable vessel segmentation. LAL mainly consists of two designs: a label adversarial loss and an embeddable adjustment layer. The label adversarial loss introduces an adversarial weight to establish adversarial between skeleton-level supervision and pixel-level supervision, while the adjustment layer adjusts the network to match the different adversarial weights. Experimental results on public datasets show that such a design can efficiently capture the variation between two supervisions, and realize the adjustment from skeleton-level segmentation to pixel-level segmentation. This process allows us to select intermediate results with clear caliber and topology, and it can also be used to generate an uncertainty map representing weak vessel boundaries and noise. The main contributions can be summarized as: (1) We implement an adjustable vessel segmentation between skeleton-level and pixel-level. (2) Our segmentation results outperform manual annotations of current public datasets and conventional filtering effects. (3) We provide an uncertainty generation method in vessel segmentation.

## 2 Label Adversarial Learning

Let us consider a typical supervised segmentation process. Given an image set X, we design a convolutional neural network $f$ and use a loss $\mathcal{L}$ (e.g., cross entropy loss) to supervise the output $y$ that approximates a given label $\tilde{y}$. When the given label $\tilde{y}$ is a skeleton-level label, the output $y$ will approximate the skeleton-level vessel segmentation, and the loss under this case can be regarded as a skeleton-level loss $\mathcal{L}_{sk}$. Similarly, when the given label $\tilde{y}$ is a pixel-level label, the loss under this case is recorded as a pixel-level loss $\mathcal{L}_p$, and the output $y$ will approximate the pixel-level vessel segmentation, and this loss can be regarded as $\mathcal{L}_p$. The process can be expressed as

$$f(X) = y \sim \mathcal{L}_{sk} \ or \ \mathcal{L}_p \tag{1}$$

We all know that the output $y$ cannot be both skeleton-level segmentation and pixel-level segmentation, thus there is a gap between the two kinds of supervision. To close the gap, we propose the label adversarial learning, which achieves an adjustable vessel segmentation between skeleton-level and pixel-level. It consists of two designs: the label adversarial loss and the adjustment layer.

### 2.1 Label Adversarial Loss

Instead of supervising with a single loss $\mathcal{L}_{sk}$ or $\mathcal{L}_p$, we build an adversarial relationship between two supervised losses via the proposed label adversarial loss:

$$\mathcal{L} = (1-w) \times \mathcal{L}_{sk} + w \times \mathcal{L}_p \tag{2}$$

where $w$ is the adversarial weight ($w \in R, 0 \leq w \leq 1$) representing the relative strength of the confrontation between the two supervisions. Different values of $w$ represent different supervision states. When $w = 0$, the loss $\mathcal{L} = \mathcal{L}_{sk}$, representing only skeleton-level label supervision. When $w = 1$, the loss $\mathcal{L} = \mathcal{L}_p$, representing only pixel-level label supervision. When $0 < w < 1$, the loss $\mathcal{L}$ represents an intermediate state from different adversarial strengths between the two supervisions. By adjusting the value of $w$, label adversarial loss $\mathcal{L}$ realizes a continuous change from skeleton-level supervision to pixel-level supervision.

### 2.2 Adjustment Layer

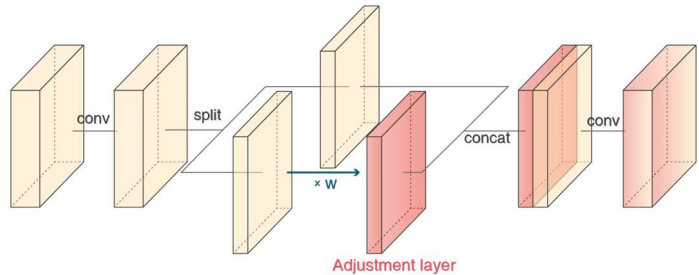

**Fig. 2.** LAL embeds the adversarial weight $w$ into CNN via the adjustment layer.

We have established a continuous process from skeleton-level supervision to pixel-level supervision via the label adversarial loss. However, if the parameters of the network are immutable, the label adversarial loss will do not work. This is because a single output can only represent one state. To obtain an adjustable continuous segmentation, we also need to add the adversarial weights $w$ into the network design, so that the network parameters can change with the adversarial weights $w$, so as to adapt to the loss $\mathcal{L}$ with different adversarial weights:

$$f(X, w) = y \sim \mathcal{L} \qquad (3)$$

We use a simple and efficient way to introduce the adversarial weights $w$ into the network as shown in Fig. 2. We multiply the adversarial weight $w$ with partial features ($0 < n < N$, $N$ is the number of channels) of the convolutional layer, and the weighted features are called the adjustment layer. After subsequent convolution, the adjustment layer will be fused with unweighted features. The proposed adjustment layer is embeddable, and we can embed it in arbitrary convolutional neural networks. In our experiments, U-Net [12] and IPN [9] were used as backbone. IPN and U-Net are two representative networks. IPN is a 3D-to-2D network whose input is a 3D OCTA volume, while U-Net is a classical 2D-to-2D network whose input is a 2D OCTA projection image. We replace the consecutive convolution in U-Net and IPN with the proposed structure in Fig. 2, and the transformed networks are called UNet-LAL (see in Fig. S1) and IPN-LAL respectively. In this way, we achieve the transformation of a static network into an adjustable dynamic network.

### 2.3    Training and Testing

We have built the network structure $f$ and the loss $\mathcal{L}$ for LAL. Next, we will introduce the training and testing process. Compared with other supervised training, LAL has one more parameter, the adversarial weight $w$. During the training process, $w$ is randomly generated floating numbers from a uniform distribution of [0,1], and the purpose is for the network to learn all possible adversarial weights. The same weight needs to be input into the network to adjust the network parameters and also needs to be fed into the loss function $\mathcal{L}$ to perform the corresponding supervision. During the testing phase, the adversarial weight $w$ becomes a freely adjustable parameter. We can adjust $w$ to achieve the change from skeleton-level segmentation to pixel-level segmentation.

## 3    Experiments

### 3.1    Datasets and Metrics

Our experiments are conducted on two recently published OCTA datasets, OCTA-500 [11] and ROSE [10]. Due to the limitations of the manual annotation in the public dataset (as shown in Fig. 6), we remake the pixel-level labels and skeleton-level labels for LAL training. We first draw pixel-level manual annotations on 100 OCTA patches (with 76 pixel × 76 pixel) from the OCTA-500 dataset. Then, they are used to train a U-Net model and test on all included images (with 304 pixel × 304 pixel) to obtain the pixel-level labels. The skeletons of these pixel-level labels are extracted using bwmorph function in MATLAB (R2021a) as skeleton-level labels. A fundus image dataset DRIVE [17] is used to verify the generality of our method.

To quantitatively analyze the continuous segmentation process of LAL, we calculated vessel diameter index (VDI), vessel density (VD), vessel length fraction (VLF) and fractal dimension (FD), which are commonly used vessel morphology metrics [13]. We also used the connectivity function introduced in [14] as vessel connectivity (VC) to measure the continuity of segmentation results. Since the vessels and capillaries are continuous structures, the isolated points in the OCTA image are often regarded as noise, so we count the number of isolated points in the binary segmentation results as the noise intensity (NI) to reflect the image noise level.

### 3.2 Skeleton-to-Pixel Adjustable Vessel Segmentation

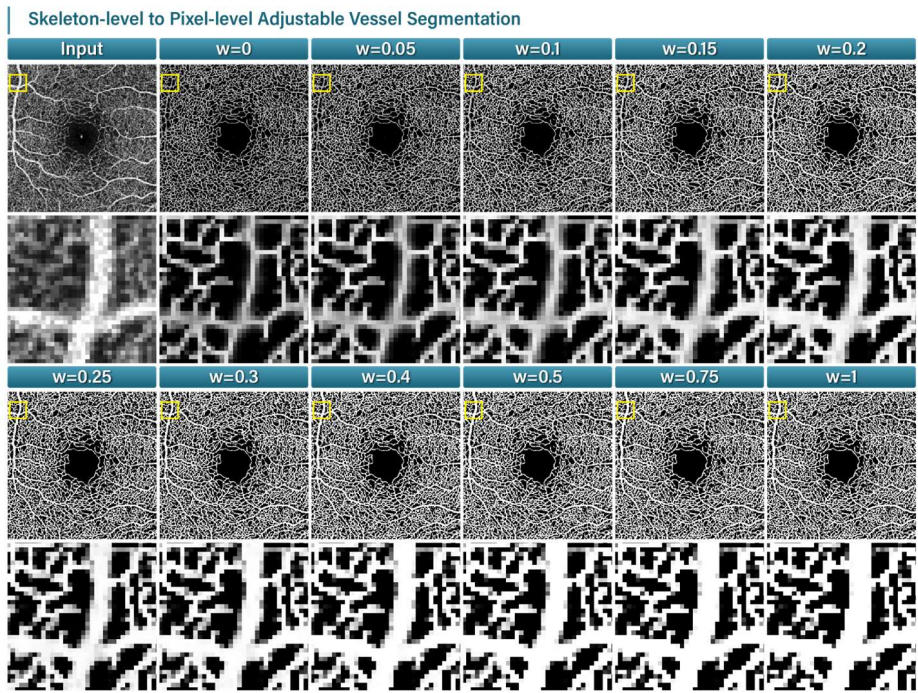

**Fig. 3.** An example of skeleton-level to pixel-level adjustable vessel segmentation. From IPN-LAL model, training and testing on OCTA-500.

We train two LAL models, IPN-LAL and UNet-LAL, on the OCTA-500 dataset. The trained model both achieve continuous changes from skeleton-level segmentation to pixel-level segmentation. Fig. 3 illustrates a typical continuous segmentation, as the adversarial weight $w$ increases, the segmentation results tend to first add the main difference on vessel diameter, and then complement some additional segmentation details such as strengthening weak boundaries, but at the same time adding some boundaries aliasing and noise.

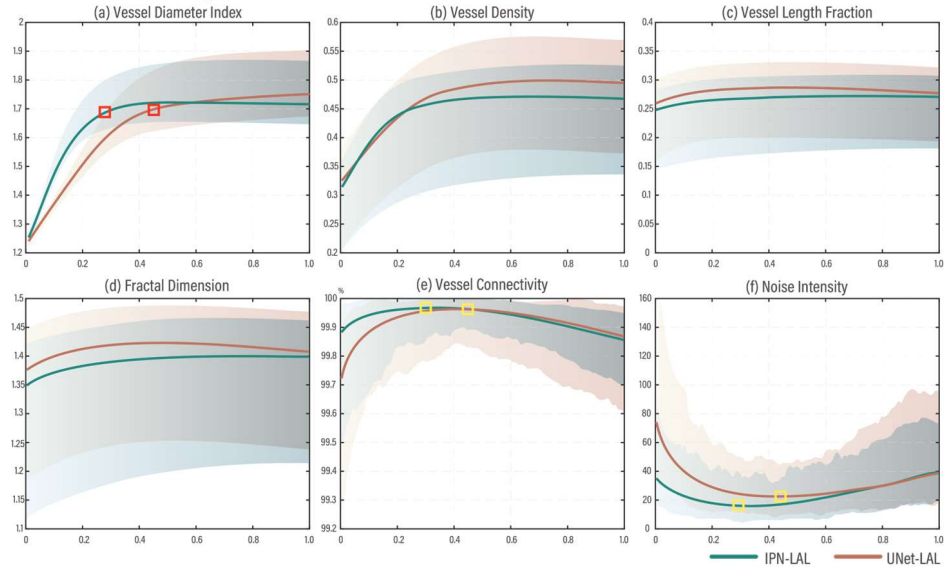

**Fig. 4.** Quantitative analysis of the changing process of the LAL segmentation results.

To quantitatively evaluate this process, we calculate the evaluation metrics of the segmentation results, as shown in Fig. 4. The results of IPN-LAL and UNet-LAL are slightly different due to their different inputs and network structures. In this paper, we are more concerned with their similar trends. VDI and VD show a trend of increasing first and then stabilizing. Significant changes in vessel diameter can be seen intuitively in the first row of Fig. 3. The changes of VLF and FD are slight. Both of them are calculated from the skeleton image, which have less difference between pixel-level labels and skeleton-level labels. VC shows a trend of increasing first and then decreasing, and NI shows a trend of decreasing first and then increasing. Trends in VC and NI show that intermediate states have better vessel continuity and less noise than boundary states.

### 3.3    Segmentation Result Recommendation

We have found that the intermediate states of the LAL results have better performance than the boundary states. The continuously adjustable segmentations of LAL allow us to choose one of the segmentation states as a recommendation result based on observations or algorithms. To recommend a result that has clear vessel diameter and clear vessel topology, we use the adversarial weight $w$ where the VDI reaches the maximum curvature, as shown in red boxes of Fig. 4(a). In this state, the VDI is close to saturation, and the VC (yellow boxes of Fig. 4(e)) and NI (yellow boxes of Fig. 4(f)) of the segmentation result are better than that of boundary states.

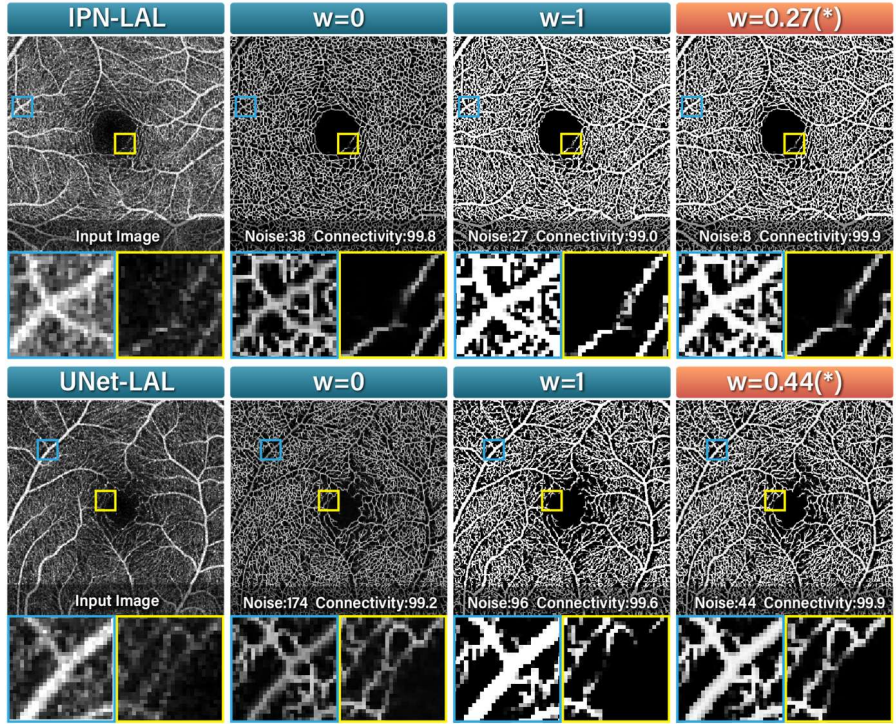

**Fig. 5.** Comparison of recommended segmentation results and boundary results for different LAL models. * represents the recommended adversarial weight.

Fig. 5 shows the differences between the recommendation results and boundary states in IPN-LAL and UNet-LAL. As shown in the blue box, the recommended results have clearer vessel caliber than the results of skeleton-level segmentation ($w = 0$). The difference in vessel diameter is enough to distinguish between arteriovenous and capillaries. As shown in the yellow box, the recommended results have a clearer vessel topology than the results of pixel-level segmentation ($w = 1$). Vessels in the recommended results are more continuous, with less noise and less aliasing. From the visual effect of the overall segmentation, the overly dense capillaries in pixel-level segmentation results make it difficult to understand, while the recommended results perform better in visual effects, which may be related to its moderate vessel density, better vessel continuity, and less noise.

### 3.4 Comparison with Manual Annotations and Filters

We compared the recommended segmentation results with manual annotations in the public dataset, OCTA-500 [11] and ROSE [10]. Fig. 6 shows that the manual annotation of OCTA-500 only contains large-caliber vessels, and the manual annotation of ROSE contains limited capillary annotation, while our recommended results segment more capillaries. This shows that our results outperform manual annotations of cur-rent public OCTA datasets. We also compared our results with the results from commonly used filters, Frangi [4] and Gabor [5]. Fig. 6 shows that the proposed method has less noise than Frangi filter and clearer vessel structures than Gabor filter. Note that, we did not compare with deep learning methods such as CS-Net [7], because the proposed LAL can embed into arbitrary convolutional neural networks.

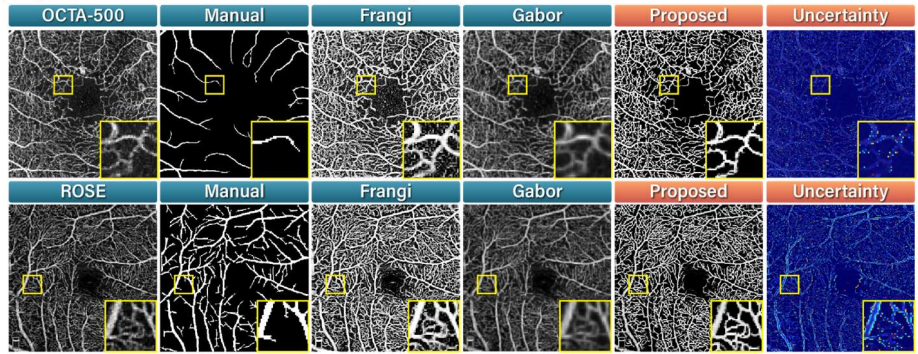

**Fig. 6.** Examples of applying our model (UNet-LAL, trained on our skeleton-level labels and pixel-level labels, $w = 0.44$) on testing images from OCTA-500 and ROSE, and the comparison with manual annotations and filters. The last column is the proposed uncertainty results.

### 3.5 Uncertainty in Vessel Segmentation

Inspired by uncertainty [15, 16], we also designed a unique uncertainty from the continuous segmentation process of the LAL. For each pixel position, we considered that the fewer times it is divided into positive samples (vessel), the higher the uncertainty. We thus calculate the probability that each position is considered to be a positive sample in a continuous process ($0 \leq w \leq 1$ with the step of 0.01). Note that, the positions that have always been considered as positive samples or negative samples need to be set to 0. As shown in Fig. 6, the uncertainty can well represent weak vessel boundaries and noise. We can further use this uncertainty to remove noise from the recommended results. Specifically, we removed points with an area less than 3 and uncertainty greater than 0.7 in the recommended results. In this way, the noise intensity of our results is further reduced, as shown in Table S1.

## 4 Conclusion

In this paper, we designed label adversarial learning, which enables adjustable vessel segmentation between skeleton-level segmentation and pixel-level segmentation. We have recommended the segmentation results with the clear caliber and topology from the continuous segmentation process. Experimental results show that the recommended results exceed the manual annotation of current public datasets and the filters. Furthermore, such a continuous process can also be used to generate an uncertainty map representing weak vessel boundaries and noise.

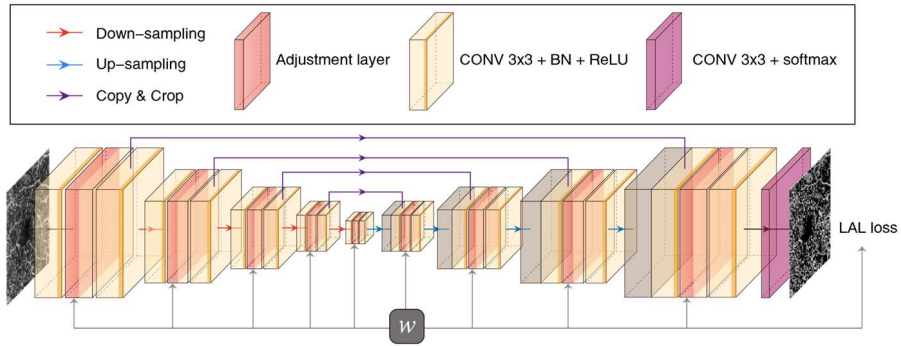

**Fig. S1.** The network architecture of UNet-LAL, which is the implementation of label adversarial learning based on the U-Net backbone.

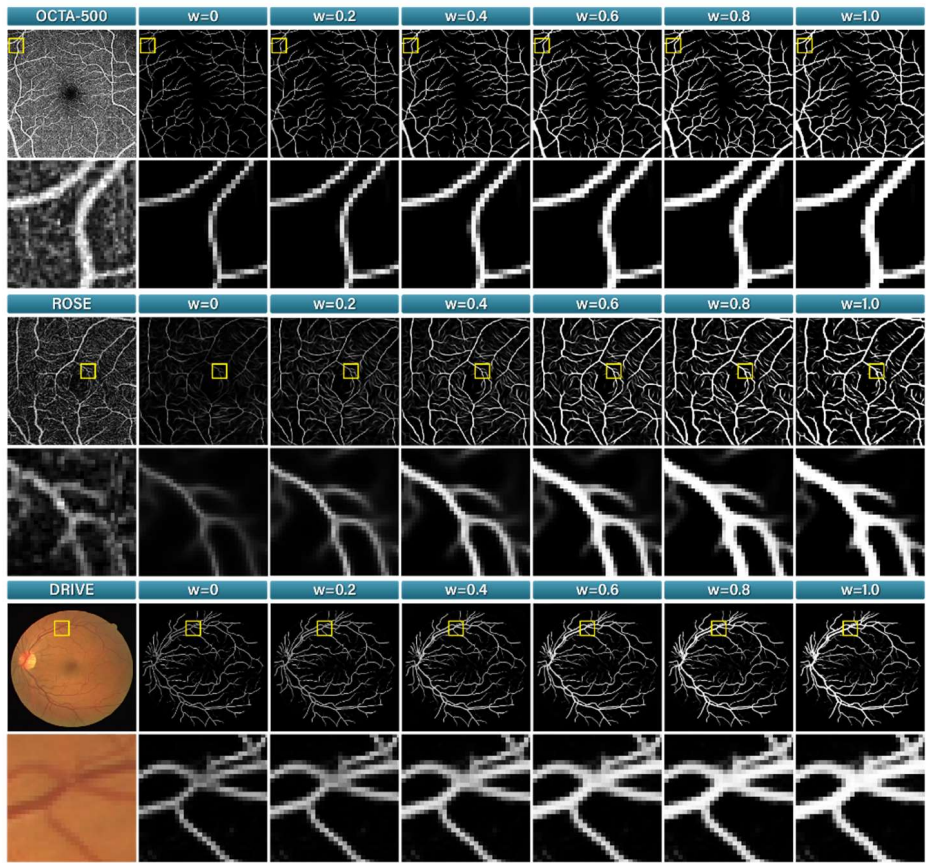

**Fig. S2.** Skeleton-level to pixel-level segmentation of Label Adversarial Learning on different datasets with their given manual annotations.

**Table S1.** Quantitative evaluation of different segmentation methods on OCTA-500 test set, using our label as the ground truth. D* represents denoising that points with high uncertainty (introduced in Sec. 3.5) are excluded.

| Methods | Accuracy | Dice | VC | NI |
|---|---|---|---|---|
| Frangi [4] | 85.5 | 86.1 | 99.3 | 194.6 |
| Gabor [5] | 67.1 | 50.7 | 98.8 | 92.1 |
| IPN-LAL | 93.9 | 93.4 | 99.9 | 16.1 |
| IPN-LAL (D*) | 93.9 | 93.4 | 99.9 | **14.5** |
| UNet-LAL | 93.4 | 93.1 | 99.9 | 21.3 |
| UNet -LAL (D*) | 93.4 | 93.1 | 99.9 | <u>14.8</u> |

Fig. S1 gives the network architecture of UNet-LAL, which is the implementation of label adversarial learning based on the U-Net [12] backbone. The implementation of IPN-LAL is similar to UNet-LAL. Fig. S2 shows that LAL can achieve skeleton-level segmentation to pixel-level segmentation on different annotations (OCTA-500 manual, ROSE manual, DRIVE manual) in different images (OCTA images, color fundus pictures). It also shows that our method can be used to adjust the vessel caliber in an elegant way (compared to corrosion and expansion). Table S1 quantitatively shows that our method is superior to the filtering method, and it also shows that our proposed uncertainty can be used to remove noise from the results, so that the noise intensity in the recommendation results is further reduced.